\documentclass[aps,prl,twocolumn,groupedaddress,floats,showpacs]{revtex4}
\usepackage{latexsym}
\usepackage{dcolumn}
\usepackage[dvips]{graphicx}
\usepackage{amssymb}
\usepackage{graphics}
\usepackage{amsmath}
\usepackage{epsf}
\usepackage{subfigure} %To include 2 figures in one column
\usepackage{dcolumn}% Align table columns on decimal point

\begin{document}
\author{E. H. Hwang and S. Das Sarma}
\title{The quasiparticle spectral function in doped graphene}
\affiliation{Condensed Matter Theory Center, Department of Physics, 
University of Maryland, College Park, MD 20742-4111}

\date{\today}

\begin{abstract}
We calculate the real and imaginary electron self-energy as well as
the quasiparticle spectral function in doped graphene taking into account
electron-electron interaction in the leading order dynamically
screened Coulomb coupling. Our theory provides the basis for
calculating {\it all} one-electron properties of extrinsic
graphene. Comparison 
with existing ARPES measurements shows broad qualitative agreement
between theory and experiment.
We also calculate the renormalized graphene
momentum distribution function, finding a typical Fermi liquid
discontinuity at $k_F$.  
We also provide a critical discussion of the relevant many body
approximations (e.g. RPA) for graphene.
\end{abstract}

\maketitle

The great deal of current activity \cite{one} in two dimensional (2D)
graphene arises from its possible technological significance as a new 2D
electronic material where carrier density can be controlled by an
external gate voltage and from its fundamental significance as a novel
2D zero band gap semiconductor system with chiral linear Dirac-like
electron-hole band energy dispersion. In particular, the chiral
linear energy dispersion with the conduction and the valence band
crossing at the ``Dirac point'' has led naturally to the interesting
analogy with 
QED whereas the gate-induced tunability of graphene carrier density
brings up the tantalizing analogy with Si MOSFETs. An important
question in this context is the extent to which the Coulomb
interaction between carriers will modify or renormalize the 
chiral linear electron-hole band dispersion in graphene. 

In this Letter we theoretically consider carrier interaction effects
in {\it extrinsic} graphene by calculating the many-body self-energy, the
quasiparticle spectral function, and the renormalized 
momentum distribution function of graphene in
the presence of free carriers (i.e. for doped or gated graphene where
carriers fill the 2D band upto a Fermi level $E_F$). Our results show
that although the 
quantitative renormalization effects of interactions on the graphene
single-particle properties are substantial, extrinsic graphene remains
an effective 2D Fermi liquid ``metal'', qualitatively preserving its
non-interacting chiral linear band dispersion even in the presence of
mutual Coulomb interaction.

The quasiparticle spectral function, $A({\bf k},\omega)$, is a central
quantity in the many-body physics of interacting systems with
$A({\bf k},\omega) \equiv -2 {\rm Im}G({\bf k},\omega)$ where $G({\bf
  k},\omega)$ is the 
single particle (retarded) Green function for momentum {\bf k} and
energy $\omega$ (we use $\hbar =1$ throughout this paper). For the
non-interacting bare system we immediately get $A_0({\bf k},\omega) = 2\pi
\delta(\omega-\varepsilon_{sk} + E_F)$ where $\varepsilon_{sk} = s
\gamma k$ with $k\equiv |{\bf k}|$ is 
the bare graphene linear band dispersion with $s=\pm 1$ denoting the
conduction ($+1$) and the valence ($-1$) band, and $\gamma \approx 10^6$
cm/s is the band velocity. We will assume that the system is n-doped
with electrons filling the graphene conduction band upto a free
carrier density ($n$) dependent chemical potential or Fermi level
given by $E_F=\gamma k_F$, where the Fermi momentum $k_F=(\pi
n)^{1/2}$. We have taken into account the spin and the valley
degeneracy of graphene in obtaining the Fermi momentum. The
noninteracting spectral function $A_0({\bf k},\omega)$ being a delta
function signifies that the band electron at momentum {\bf k} has
{\it all} its spectral weight precisely at the energy
$\varepsilon_k=\gamma k$, 
i.e. the noninteracting particle exists {\it entirely} at the energy
$\gamma k$ for a given momentum {\bf k}. In the presence of
interaction effects, the many-body self-energy function
$\Sigma({\bf k},\omega)$ modifies the single-particle Green function:
$G^{-1}({\bf k},\omega) = G_0^{-1}({\bf k},\omega)-\Sigma({\bf
  k},\omega)$, and the 
corresponding interacting or renormalized spectral function is given
by 
\begin{equation}
A({\bf k},\omega) \equiv \frac{2 {\rm Im}\Sigma({\bf k},\omega)}{\left
    [\omega-\xi_k-{\rm Re}\Sigma({\bf k},\omega)\right ]^2+\left [ {\rm
      Im}\Sigma({\bf k},\omega)\right ]^2} 
\end{equation}
where $\Sigma({\bf k},\omega)={\rm Re}\Sigma({\bf k},\omega) + i {\rm
  Im}\Sigma({\bf k},\omega)$ is complex, and $\xi_{sk} \equiv
\varepsilon_{sk}-E_F$. In general, $A({\bf k},\omega)$ could be a
complicated function of ${\bf k}$ and $\omega$, and there is no guarantee
that it will have a delta function peak defining a
quasiparticle. We note that 
$\int \frac{d\omega}{2\pi}A({\bf k},\omega)=1$ is a sum rule, guaranteeing
that the electron at momentum {\bf k} exists in the whole energy
space, but it may exist completely incoherently spread out over the
whole $\omega$-space without any coherent structure (i.e. a delta
function at $k=k_F$), indicating a complete failure of the Fermi
liquid picture. 
%where the interacting system is taken to be a collection of
%quasiparticles with well-defined energy at the Fermi momentum
%$k_F$. 
Thus, the Fermi liquid theory applies {\it only when} the
renormalized spectral function $A(k=k_F,\omega)$ at the Fermi momentum
has a delta function peak, i.e. $A(k_F,\omega)$ can be written
as  $A(k_F,\omega)=2\pi Z \delta(w-\xi_k^*) + A_{\rm in}(\omega)$,
where $Z$ is the so-called ``renormalization factor'', $\xi_k^*$
denotes the renormailzed quasiparticle energy (measured from the chemical
potential), and $A_{\rm in}$
is the incoherent background spectral function. If $A(k_F,\omega)$
does {\it not} have any delta function peak 
at all, then the system is a
non-Fermi liquid. 

\begin{figure}
\bigskip
\epsfxsize=0.81\hsize
\hspace{0.0\hsize}
\epsffile{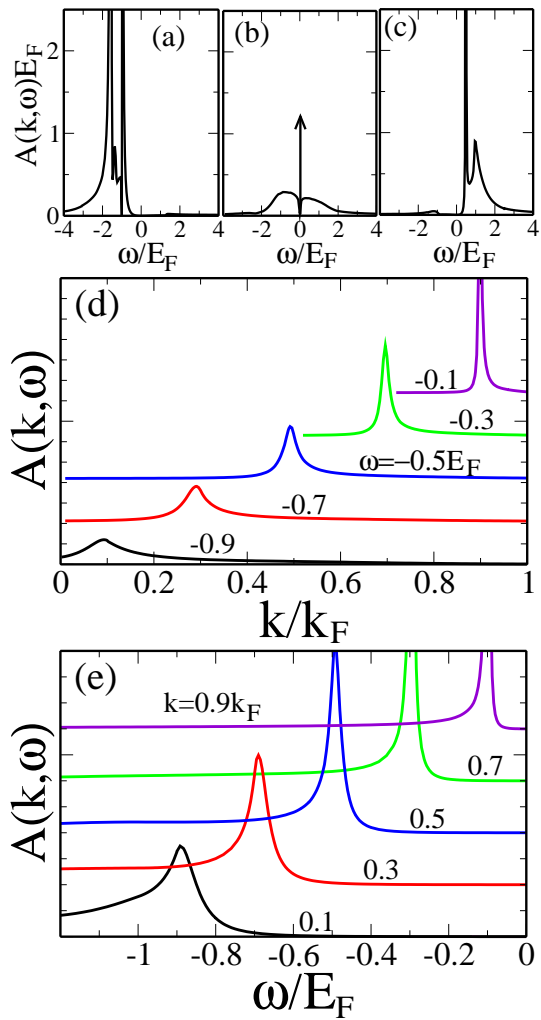}
\epsfxsize=0.81\hsize
\epsffile{fig_1d.eps}
\epsfxsize=0.81\hsize
\epsffile{fig_1e.eps}
\caption{\label{Fig1} 
Calculated graphene spectral function for different wave vectors
(a) $k=0$, (b) $k=k_F$, (c) $k=1.5 k_F$ without any disorder effects. (d)
Spectral function as a function 
of wave vector for different energies (MDC) and
(e) as a function of energy for different
wave vectors (EDC). 
In (d) and (e) we include an impurity scattering rate of $0.5E_F$ as
explained in Fig. 3(b). We only show the spectral features above the
Dirac point.
Here all energies are measured from
the Fermi level, and the Dirac point is at $\omega = -E_F$. 
}
\end{figure}

In Fig. 1 we show our calculated interacting quasiparticle spectral
function for extrinsic graphene at a fixed carrier density $n=10^{12}$
cm$^{-2}$. 
%corresponding to $E_F= 116$ meV. 
The
calculations are carried out at $T=0$, but thermal effects are
unimportant here since $T/T_F \ll 1$ even at room temperatures. The
following salient features of graphene quasiparticle spectral function
are notable in Fig.1: (1) There is a well-defined delta-function
quasiparticle peak at $k_F$ with a rather substantial spectral weight
of $Z \approx 0.85$, indicating that  15\% of the bare
spectral weight goes into incoherent background.
(2) The quasiparticle spectral function for $k \neq k_F$ shows,
in general, broadened peak structures indicating damped
quasiparticles. 
(3) The generic broadened double-peak structure at $k \neq
k_F$ indicates that 
away from the Fermi surface, the renormalized
graphene spectra would have
%show strong qualitative deviation from the
%noninteracting $\varepsilon_k=\gamma k$ type behavior with there being 
two distinct energies 
%at each $k$ ($\neq k_F$) value showing strong spectral (damped) peaks 
--- the second peak, which has been
well-studied in the literature in both 2D \cite{Jalabert} and 3D
\cite{Hedin} interacting electron systems, is often referred to as the
``plasmaron'' peak, indicating a coupled electron-plasmon composite
excitation. 
(4) All spectral functions for $k\neq k_F$ have finite
width corresponding to quasiparticle damping defined by the imaginary
part of the many-body self-energy, Im$\Sigma(k,\omega)$.
Note that in Figs. 1(d) and (e) disorder contributes to combining the
double peak structure into a 
very broadened single peak.

Since Fig.\;1 giving the interacting graphene quasiparticle spectral
function is the central result being presented in this paper, we first
discuss the importance and the implications of our calculated spectral
function before describing the details of our theory and other
results. First, 
%contrary to other claims in the literature, 
extrinsic
(i.e. gated or doped) graphene 
%in the presence of free carriers
is a Fermi
liquid with a well-defined undamped quasiparticle at $k_F$.
Thus, the chiral Dirac-like
linear dispersion of graphene band structure does not lead to any
anomalous non-Fermi liquid behavior. Second, extrinsic graphene has
well-defined, but damped, quasiparticle peaks for all
momenta. One direct experimental probe of the quasiparticle
spectral function is the tunneling spectroscopy, which has
been studied extensively in GaAs-based 2D systems \cite{Eisenstein},
but has not yet been studied in graphene. The measurement that comes
closest to studying $A({\bf k},\omega)$ in graphene is the
angle-resolved-photo-emission spectroscopy (ARPES)
\cite{Rotenberg,Lanzara}. Unfortunately, there are problems in
directly comparing our theoretical spectral function of Fig.\;1 
with the experimental ARPES results. One problem is that
electron-phonon interaction induced many-body renormalization
\cite{phonon} and disorder also contribute to the graphene spectral
function.  
In Figs. 1(d) and (e), we show the theoretical results corresponding to the
experimental ARPES spectra --- in particular, Figs. 1(d) and 1(e)
respectively correspond to
the so called MDC (momentum distribution curves) and EDC (energy
distribution curves) of ARPES data.
%But, the more serious problem is the ARPES is {\it not} a direct
%measurement of $A(k,\omega)$, but actually a measurement of momentum
%or energy averaged spectral function (the so-called EDC and MDC
%respectively, the energy- and momentum- distribution curves). 
Since
the actual theoretical spectral function, Fig. 1, has very strong
momentum ($k$) and energy ($\omega$) dependence, a direct comparison
with ARPES data will necessarily involve detailed instrumental issues
involving resolution, the ($k$,$\omega$) regime of averaging, and
instrumental errors which are all well 
beyond the scope of the current theoretical work. A very cursory
comparison between our results and the experimental ARPES data
\cite{Rotenberg,Lanzara}
show reasonable qualitative
agreement, but short of large-scale data-fitting, one cannot 
make definitive quantitative statements.
One important point to note here is that the plasmaron structure
(i.e. the additional peak) does not really show up in the MDC and EDC
spectra since they carry small spectral weight compared with the main
quasiparticle peak, particularly in the energy regime near $E_F$ as
observed experimentally \cite{Rotenberg}.

We now describe the theory leading to our calculation of the
interacting quasiparticle spectral function depicted in Fig. 1. The
self-energy, $\Sigma({\bf k},\omega)$, defining the spectral function
through Eq. (1) is given, in the leading order dynamically screened
Coulomb interaction approximation\cite{Hedin}
% by a sum of an exchange and a
%correlation part:
\begin{eqnarray}
\Sigma_s({\bf k},i\omega_n) =
-k_BT\sum_{s'}\sum_{{\bf q},i\nu_n} 
G_{0,s'}({\bf k}+{\bf q},i\omega_n+i\nu_n) \nonumber \\
\times \frac{V_c({\bf q})}{\epsilon(q,i\nu_n)}
F_{ss'}({\bf k},{\bf k}+{\bf q}),
\end{eqnarray}
where $V_c(q)=2\pi e^2/\kappa q$ is the 2D Coulomb interaction with
the background lattice dielectric constant $\kappa$ and 
$F_{ss'}({\bf k},{\bf k}') = (1 + ss' \cos\theta_{{\bf
kk}'})/{2}$ is a sublattice overlap matrix element arising from
graphene band structure, and
$\epsilon(q,\omega)=1+V_c(q)\Pi(q,\omega)$ is the dynamical RPA
dielectric function for graphene with the irreducible electron-hole
polarizability 
$\Pi(q,\omega)$, which has recently been calculated
\cite{rpa}. We note that the
irreducible self-energy approximation used in our theory corresponds
to keeping the Coulomb interaction and the infinite series of
electron-hole polarizability bubble diagrams in the self-energy calculation.

An important technical point in the calculation of the self-energy is
that, in general, two distinct types of field-theoretical divergence
may appear in the theory: infrared (small momentum) and ultraviolet
(large momentum). The infrared divergence 
arises from the $1/q$ long-range ($q\rightarrow 0$) divergence of the
Coulomb interaction, and is regularized by our RPA theory through
screening, i.e. the fact that $V_c(q)/\epsilon(q,\omega)$ does not
have a $q\rightarrow 0$ infrared divergence. While the infrared
self-energy divergence, which is regularized by RPA screening, is
generic to all problems involving long-range Coulomb interaction, the
ultraviolet divergence, which arises from the peculiar graphene band
dispersion, is specific to graphene in the sense that it
does not occur in the corresponding parabolic band 2D (or 3D) system
\cite{Jalabert,Hedin}. The ultraviolet divergence in the graphene
self-energy is fixed by realizing that the linear Dirac dispersion of
graphene only applies upto momenta of the order of inverse lattice
constant, and therefore all momentum integrals should have an upper
cut off $k_c \sim 1/a$, where $a$ is the graphene lattice constant. We
emphasize that there is nothing {\it ad hoc} or mysterious about this
ultraviolet regularization since the band dispersion in graphene would
deviate strongly from linearity for $k\ge k_c$. We also note that the
infrared divergence associated with the $q\rightarrow 0$ singularity of
the long-range Coulomb interaction and ultraviolet divergence
associated with the $q > k_c \sim a^{-1}$ lattice wave vector scales
have completely different origins, and while RPA screening fixes the
infra-red divergence, the momentum cut-off fixes the ultraviolet
divergence. Both infra-red and ultra-violet regularizations are
important for graphene in contrast to 2D (or 3D) parabolic dispersion
interacting electron systems where only the infra-red divergence is
germane.

\begin{figure}
\bigskip
\epsfxsize=0.7\hsize
\hspace{0.0\hsize}
\epsffile{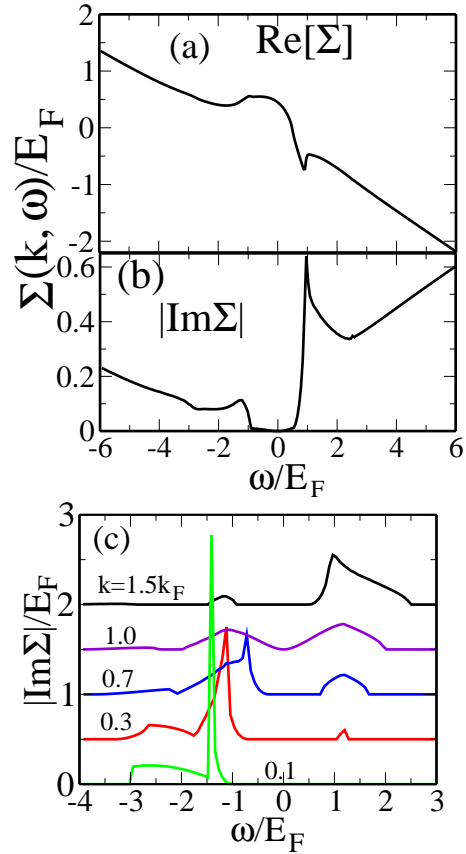}
\epsfxsize=0.65\hsize
\epsffile{fig_2b.eps}
\caption{\label{Fig2} Calculated graphene self energy for $k=1.5k_F$
  as a function of energy, (a) Re$\Sigma$ and
  (b) $|{\rm Im}\Sigma|$ of correlation part. (c) 
  Plasmon contribution to the imaginary 
  part of self energy for several wave vectors. The Dirac point is at
  $\omega = -E_F$.
}
\end{figure}

In Fig. 2 we show our calculated graphene self-energy for $n=10^{12}$
cm$^{-2}$ at $k=1.5k_F$. We note that 
the peak in Im$\Sigma$ correlates with the dip in the Re$\Sigma$, and
most of the spectral weight resides near dip of Re$\Sigma$
(or equivalently, at the peak of Im$\Sigma$). We emphasize that, as is
apparent from the detailed spectral function in Fig. 1, the
various structures in the self-energy lead to specific features in
$A(k,\omega)$ --- in particular, Re$\Sigma$ and Im$\Sigma$ control
respectively the energy renormalization and the broadening (or
damping) of the quasiparticle. An important issue in this context is
the precise contribution of the collective plasmon mode to Im$\Sigma$,
which we have evaluated explicitly as shown in Fig. 2(c). Only far
away from the 
Fermi surface the plasmon contributions are strong, giving rise
to the strong plasmaron peak in the spectral function.

In Fig.\;3 we show our calculated renormalized quasiparticle momentum
distribution function 
$n(k) = \int_{-\infty}^{E_F} \frac{d\omega}{2\pi} A(k,\omega)$
for extrinsic graphene and disorder effects on the spectral
function. The  effect of 
electron-electron interaction on $n(k)$ is rather obvious in
Fig. 3(a). We note 
that the interacting momentum distribution function has a
discontinuity of relative size $Z$ at $k=k_F$, clearly
establishing the Fermi liquid behavior of extrinsic (i.e. doped or
gated) graphene. The discontinuity increases as $r_s$ decreases.
We note here that the corresponding intrinsic (or undoped) graphene
does not have a discontinuity in the distribution function since it is
a marginal Fermi liquid \cite{Dassarma}.

Impurity effects are usually introduced diagrammatically into the
RPA screening  by including ladder impurity diagrams into the
electron-hole bubble. We use a particle-conserving
expression\cite{mermin}, which captures the essential physics of
impurity scattering effects on the electron polarizability $\Pi$.
We find that disorder strongly
suppresses plasmaron peak, but not the
quasiparticle peak. Thus the double peak structure in the spectral
function becomes a very broadened single peak. 
In the strongly disordered experimental graphene samples therefore
we only have a
broadened quasiparticle peak.

\begin{figure}
\bigskip
\epsfxsize=1.\hsize
\hspace{0.0\hsize}
\epsffile{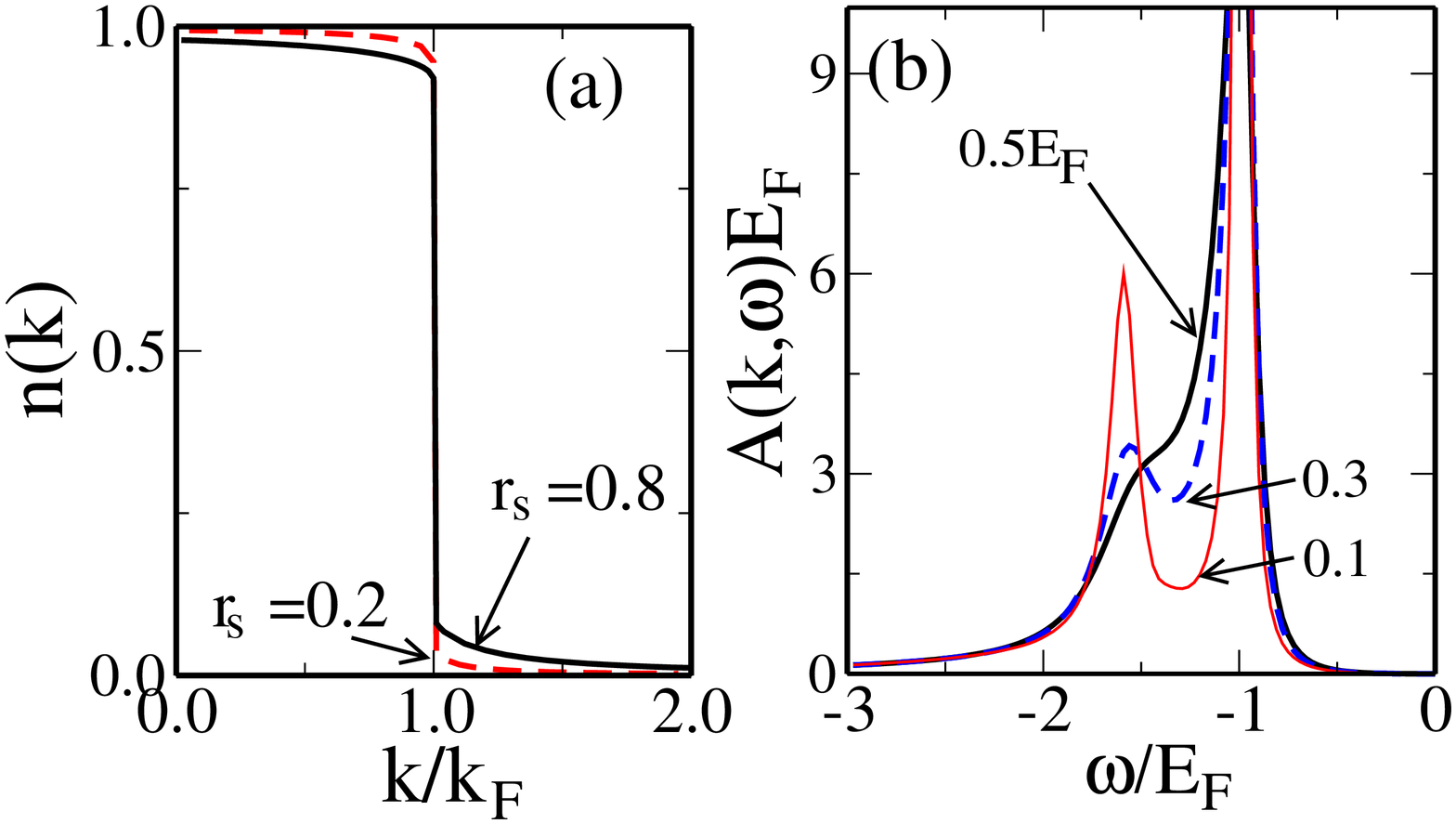}
\caption{
\label{Fig3} 
(a) Renormalized momentum distribution function as a function of momentum
for $r_s=0.8$ (solid line) and 0.2 (dashed line). (b) Disorder effects
on the $k=0$ spectral function for different impurity scattering rates of
0.1, 0.3, 0.5$E_F$. 
}
\end{figure}

We now comment on the experimental implications, the theoretical
approximations, and the connections to  earlier work. As
mentioned before, the graphene ARPES measurements have information on
the electron spectral function, which are, however, complicated by the
presence of additional interaction effects such as electron-impurity
(i.e. disorder) and electron-phonon interactions. If these two
interaction effects can be subtracted out (e.g. $\sim$ the 200 meV
structure \cite{Rotenberg}
presumably arising from phonons in the experimental data), then ARPES
measurements could indeed be compared with our calculated spectral
function shown in Fig.1. 
%In particular the ARPES MDC results should be
%compared with our Fig. 1(d) where $A(k,\omega)$ is plotted as a
%function of momentum $k$ for different fixed values of energy $\omega$
%showing the (damped) quasiparticle peaks. 
%Note that any negative
%energy structure residing below the Dirac point (at $\omega=0$) would
%not be observable in the experiment, and therefore the interesting
%double-peak spectral features, the quasiparticle peak and the
%plasmaron peak, so prominent for $k\neq k_F$ in Fig. 1(a) and(c) would
%not show up in the experimental ARPES data, and, as our Fig. 2(d)
%shows, plasmon contributions to the quasiparticle spectral weight are
%small except at large momentum (e.g. $k=1.5k_F$). 
%Similarly, the ARPES
%EDC spectra should mimic our Fig. 1(e) where $A(k,\omega)$ is plotted
%as a function of $\omega$ for different momentum. 
We note that, in
agreement with the ARPES data, the quasiparticle spectral peak becomes
wider due to plasmon contribution as the energy approaches to the Dirac
point ($\omega=-E_F$).
%only one broadened spectral peak shows
%up for $k \neq k_F$ for $\omega > -E_F$ (i.e. above the Dirac point)
%except some small plasmaron structures at large momenta which would
%probably be unobservable due to impurity and phonon effects.
Including disorder effects, indeed there is good agreement between
theory and experiment.

Our theory is based on the leading-order expansion of the electron
self-energy using the dynamically screened Coulomb interaction (the
so-called infinite bubble diagram expansion with each bubble being the
non-interacting electron-hole polarizability \cite{rpa}), which 
we believe to be a quantitatively accurate approximation for extrinsic
graphene (i.e. at any finite carrier density)  by
virtue of graphene having a reasonably small (and density independent)
dimensionless interaction parameter $r_s=e^2/\kappa \gamma \sim 0.8$ (0.4)
for graphene on SiO$_2$ (SiC) substrate. 
%The infra-red
%divergence of the Coulomb interaction is well-captured by the infinite
%bubble diagram expansion approximation of our theory, and we believe
%that, except for the ultraviolet divergence associated with the lattice
%cut-off, our theory is well-controlled in extrinsic graphene with its
%relatively weak ($r_s <1$) Coulomb coupling. 
In fact, the same
self-energy approximation, often referred to in the literature as
``RPA'' or ``GW'' approximation, is known to work well in 3D alkali
metals where $r_s \sim 3-6$ \cite{Hedin} and in 2D semiconductor system
\cite{Jalabert,Eisenstein} where $r_s \sim 1-10$. 
For extrinsic graphene only the terms having
bubble diagrams (or RPA type diagrams) in self-energy  have the infra-red
divergence. All other vertex corrections (i.e. non-RPA type diagrams)
have finite values. Thus, most contributions (for $r_s <1$) come from
RPA type diagrams. Since we can
control the ultra-violet divergence by introducing a physical energy
cut-off, our RPA approach is quantitatively accurate for
extrinsic graphene.   
However, for intrinsic graphene, there is no infra-red divergence in
RPA type diagrams, only the ultra-violet
divergence. Since all diagrams then contribute to the self-energy, the
perturbative expansion  does not converge
for intrinsic graphene. Therefore, RPA is not a meaningful
approximation for intrinsic graphene, and indeed it predicts a non-Fermi
liquid behavior \cite{Dassarma}.

In discussing
connection between our work and earlier work, we  mention that,
although there has been a great deal of recent theoretical work on 
interaction effects in graphene, much of it has focused on intrinsic
graphehe \cite{new}, and most of the work on extrinsic graphene
has focused on thermodynamic properties \cite{exchange}. 
%To the best
%of our knowledge, ours is 
%the first calculation of the quasiparticle spectral function in
%extrinsic graphene in the presence of electron-electron interaction. 
A recent calculation of the quasiparticle spectral function in graphene
due to electron-phonon interaction has appeared \cite{phonon} in the
literature, and within the weak-coupling theory (i.e. both
electron-electron and electron-phonon interactions being weak), the
total spectral function should be a sum of these two spectral
functions. We do note that our calculated spectral function and
self-energy in extrinsic graphene is qualitatively rather similar to
that in ordinary parabolic band 2D carrier systems \cite{Jalabert},
thus pointing to the fact that, in spite of its chiral Dirac-like band
dispersion, doped graphene in the presence of free carriers
is qualitatively similar to doped 2D semiconductor systems.

In summary, we have calculated the electron spectral function in 2D doped
(i.e. extrinsic) graphene, finding our theory to be in good
qualitative agreement with the available experimental data. Our theory
convincingly demonstrates the Fermi liquid nature of doped graphene.

This work is supported by U.S. ONR and LPS-NSA.


\begin{thebibliography}{14}
\expandafter\ifx\csname natexlab\endcsname\relax\def\natexlab#1{#1}\fi
\expandafter\ifx\csname bibnamefont\endcsname\relax
  \def\bibnamefont#1{#1}\fi
\expandafter\ifx\csname bibfnamefont\endcsname\relax
  \def\bibfnamefont#1{#1}\fi
\expandafter\ifx\csname citenamefont\endcsname\relax
  \def\citenamefont#1{#1}\fi
\expandafter\ifx\csname url\endcsname\relax
  \def\url#1{\texttt{#1}}\fi
\expandafter\ifx\csname urlprefix\endcsname\relax\def\urlprefix{URL }\fi
\providecommand{\bibinfo}[2]{#2}
\providecommand{\eprint}[2][]{\url{#2}}


\bibitem[{\citenamefont{\mbox{Das Sarma} et~al.}(2007)\citenamefont{\mbox{Das
  Sarma}, Geim, Kim, and MacDonald}}]{one}
\bibinfo{editor}{\bibfnamefont{S.}~\bibnamefont{\mbox{Das Sarma}}},
  \bibinfo{editor}{\bibfnamefont{A.~K.} \bibnamefont{Geim}},
  \bibinfo{editor}{\bibfnamefont{P.}~\bibnamefont{Kim}}, \bibnamefont{and}
  \bibinfo{editor}{\bibfnamefont{A.~H.} \bibnamefont{MacDonald}}, eds.,
  \emph{\bibinfo{title}{Exploring Graphene: Recent Research Advances, A Special
  Issue of Solid State Communications}}, vol. \bibinfo{volume}{143}
  (\bibinfo{publisher}{Elsevier}, \bibinfo{year}{2007}); references therein.

\bibitem{Jalabert} R. Jalabert and S. Das Sarma, \prb {\bf 40}, 9723
  (1989).

\bibitem{Hedin} L. Hedin and S. Lundqvist, in {\it Solid State
    Physics}, edited by H. Ehrenreich {\it et al.} (Academic, New
  York, 1969), Vol. 23.

\bibitem{Eisenstein}S. Q. Murphy {\it et al}., 
%J. P. Eisenstein, L. N. Pfeiffer,
%  and K. W. West, 
Phys. Rev. B {\bf 52}, 14825 (1995);
J.P. Eisenstein {\it et al.}, 
%D. Syphers, L.N. Pfeiffer, and K.W. West, 
Solid  State Commun. {\bf 143}, 365 (2007).

\bibitem{Rotenberg} A. Bostwick {\it et al}., Nature Phys. {\bf 3}, 36
  (2007). 

\bibitem{Lanzara} S. Y. Zhou {\it et al.}, Nature Phys. {\bf 2}, 595
  (2006). 

\bibitem{phonon} M. Calandra and F. Mauri, arXiv:0707.1467; C. H. Park 
  {\it et al}., arXiv:0707.1666; W. K. Tse and S. Das Sarma,
  arXiv:0707.3651. 

\bibitem{rpa} E. H. Hwang and S. Das Sarma, Phys. Rev. B {\bf 75},
  205418 (2007).

\bibitem{Dassarma}S. Das Sarma {\it et al}., Phys. Rev. B {\bf 75},
  121406 (2007).

\bibitem{mermin}D. Mermin, \prb {\bf 1}, 2362 (1970).

\bibitem{new}] J. Gonz\'{a}lez {\it et al}. Phys. Rev. Lett. {\bf 77},
  3589 (1996);  D. V. Khveshchenko, Phys. Rev. B {\bf 74}, 161402(R) (2006).



\bibitem{exchange} E. H. Hwang {\it et al.}, 
%B. Y. K. Hu, and S. Das Sarma,
  cond-mat/0703499;
% E. H. Hwang {\it et al.}, 
%B. Y. K. Hu, and S. Das Sarma,
  cond-mat/0612345;
Y. Barlas {\it et al.}, Phys. Rev. Lett. {\bf 98}, 
  236601(2007); M. Polini {\it et al}., Solid. State. Commun. {\bf
    143}, 58 (2007). 


%\bibitem{Ando_RMP} T. Ando {\it et al}., Rev. Mod. Phys. {\bf 54}, 437
%  (1982). 



\end{thebibliography}
\end{document}